\documentclass[12pt,preprint]{aastex}
\usepackage{psfig}
\usepackage{graphicx}

\newcommand{\nswd}{NS$-$WD}

\newcommand{\tlife}{$\tau_{\rm life}$}
\newcommand{\tmrg}{$\tau_{\rm mrg}$}
\newcommand{\td}{$\tau_{\rm d}$}
\newcommand{\tc}{$\tau_{\rm c}$}
\newcommand{\tsd}{$\tau_{\rm sd}$}
\newcommand{\rate}{$\cal R$}
\newcommand{\rtot}{${\cal R}_{\rm tot}$}

\newcommand{\rpeak}{${\cal R}_{\rm peak}$}

\newcommand{\ntot}{$N_{\rm tot}$}

\newcommand{\nobs}{$N_{\rm obs}$}
\newcommand{\nmean}{$<$$N_{\rm obs}$$>$}
\newcommand{\solarM}{M$_{\rm \odot}$}
\newcommand{\chmass}{$\cal M$}

\shorttitle{COALESCENCE RATE ESTIMATES II. NS--WD.}
\shortauthors{C. Kim et al.}

\begin{document}

\title{THE PROBABILITY DISTRIBUTION OF BINARY PULSAR COALESCENCE RATE
ESTIMATES. II.~NEUTRON STAR--WHITE DWARF BINARIES.}

\author{CHUNGLEE KIM\altaffilmark{1}, VASSILIKI
KALOGERA\altaffilmark{1}, DUNCAN R. LORIMER\altaffilmark{2}, AND
TIFFANY WHITE\altaffilmark{1}}

\affil{ $^{1}$ Northwestern University, Dept. of Physics \& Astronomy,
2145 Sheridan Road, Evanston, IL 60208\\ 
$^{2}$ University of Manchester, Jodrell Bank Observatory, Macclesfield, 
Cheshire, SK11 9DL, UK.\\ 
c-kim1@northwestern.edu; vicky@northwestern.edu;
drl@jb.man.ac.uk; tiphane3@hotmail.com}

\begin{abstract}
We consider the statistics of pulsar binaries with white
dwarf companions (\nswd). Using the
statistical analysis method developed by Kim et al.\ (2003) we
calculate the Galactic coalescence rate of \nswd~binaries due to
gravitational-wave emission. We find that the most likely values for
the total Galactic coalescence rate (\rtot) of \nswd~binaries lie in
the range 0.2--10 Myr$^{-1}$ depending on different assumed pulsar
population models. For our reference model, we obtain 
${\cal R}_{\rm tot}=4.11_{-2.56}^{+5.25}$ Myr$^{-1}$ at a 68\% 
statistical confidence level. These rate estimates are not corrected 
for pulsar beaming and as such they are found to be about a factor of 
20 smaller than the Galactic coalescence rate estimates for double 
neutron star systems.
Based on our rate estimates, we calculate the gravitational-wave
background due to coalescing \nswd~binaries out to extragalactic distances
within the frequency band of the Laser Interferometer Space Antenna.
We find the contribution from \nswd~binaries to 
the gravitational-wave background to be negligible.
\end{abstract}


\section{INTRODUCTION}
\label{sec:intro}

The observed properties of double neutron star (DNS) systems along
with models of pulsar survey selection effects have been used for many
years in order to estimate the coalescence rate of DNS systems due to
the emission of gravitational radiation {\citep{nps91,ph91,cl95}}. In
Kim, Kalogera, \& Lorimer (2003; hereafter paper I\nocite{kkl03}), 
we presented a novel method to calculate 
the probability distribution of the coalescence
rate estimates for pulsar binaries (see also \citet{k04}). 
In paper I, we applied this method to Galactic DNS
systems. Having a probability distribution at hand allowed us to
calculate the most likely value for the Galactic DNS coalescence rate
as well as statistical confidence limits associated with it. These
were then used to calculate the expected DNS inspiral detection rate
for Laser Interferometer Gravitational-Wave Observatory
{\citep{ligo92}}. In this second paper, we extend our study to
\nswd~binaries that are relevant to the future NASA/ESA mission 
Laser Interferometer Space Antenna (LISA; Bender et al.\ 1998{\nocite{lisa98}}) . 

There are now more than 40 neutron star-white dwarf (\nswd) binary
systems known in the Galactic disk (see e.g.~{\citet{d01}} for a review). 
Here, we consider the subset of \nswd~binaries which will coalesce due to
gravitational-wave (GW) emission within a Hubble time.
There are currently three such {\it coalescing} binaries known:
PSR~J0751+1807 {\citep{lzc95}}, PSR~J1757$-$5322 {\citep{eb01}}, and
PSR~J1141$-$6545 {\citep{k00,b03}}. We calculate the Galactic
coalescence rate of \nswd~binaries based on their observed properties
using the method introduced in paper I. The GW frequencies emitted by
these systems fall within the $\sim0.1-100$ mHz frequency band of the
LISA{\nocite{lisa98}}. Using our
rate estimates we calculate the GW amplitude due to the \nswd~binaries
out to cosmological distances and compare it to the sensitivity curve of LISA
{\citep{lisa00}} as well as the Galactic confusion noise estimates from
white dwarf binaries {\citep{bh90,bh97,nyp01,s01}}.

The organization of the rest of this paper is as follows.  In \S
\ref{sec:coal}, we consider the lifetimes of \nswd~binaries and summarize the
techniques we use to calculate their coalescence rate. 
The results of these calculations are presented in \S \ref{sec:results}.
In \S \ref{sec:lisa} we use our rate results to calculate
the expected GW backgound produced by \nswd~binaries
Finally, in \S \ref{sec:discussion},
we discuss the implications of our results.

\section{COALESCING \nswd~BINARIES}\label{sec:coal}

In general, the coalescence rate of a binary system containing an
observable radio pulsar is defined by
\begin{equation}
{\cal R} = {\frac{N_{\rm PSR}}{\tau_{\rm life}}} \times {f_{\rm b}}~,
\end{equation}
where $N_{\rm {PSR}}$\footnote{This is equivalent to the so-called
`scale factor' {\citep{n87}}.} is the estimated number of pulsars in
our Galaxy with pulse profiles and orbital characteristics similar to
those of the known systems, $f_{\rm b}$ is a correction factor for
pulsar beaming and $\tau_{\rm life}$ is the lifetime of the binary
system. In the following subsections, we calculate the total lifetime
of a pulsar binary and derive the probability density
function (PDF) of the Galactic coalescence rate, $P$(\rate), for
\nswd~binaries. 

\subsection{Lifetime of a \nswd~binary}

In Table \ref{tab:obsprop}, we summarize the observational properties
and relevant lifetimes for the 3 pulsar systems considered in this
work. We define the lifetime of a coalescing pulsar binary $\tau_{\rm
life}$ to be the sum of the current age of the observable pulsar and
the remaining lifetime of the system. Assuming the pulsar spins down
from an initial period $P_0$ to the currently observed value $P$ (both
in s) due to a non-decaying magnetic dipole radiation torque (see
e.g.~{\citet{mt77}}), its current ``spin-down'' age
\begin{equation}\label{eq:tc}
\tau_{\rm sd}=\frac{P}{2\dot{P}} \left(1-\left[\frac{P_0}{P}\right]^2\right),
\end{equation}
where $\dot{P}$ (in ss$^{-1}$) is the observed period derivative. For young pulsars
like J1141$-$6545, it is usually assumed that $P_0 \ll P$ so that
$\tau_{\rm sd}$ reduces to the familiar characteristic age $\tau_{\rm
c}=P/(2 \dot{P}) \simeq 1.5$ Myr. For the older recycled pulsars, however,
{\citet{acw99}} pointed out that this assumption is usually not appropriate, 
since the weaker magnetic fields of these objects
mean that their present spin periods are only moderately larger than
the periods produced during accretion. Adopting the spin-up line from
{\citet{acw99}}, we may write
\begin{equation}\label{eq:sp}
  P_0 = \left(\frac{\dot{P}}{1.1 \times 10^{-15}}\right)^{3/4} \, {\rm s}~.
\end{equation}
Using the above two equations we calculate $\tau_{\rm sd}$ for the two
recycled pulsars J0751+1807 and J1757$-$5322 to be 6.7 and 4.9 Gyr
respectively (see Table \ref{tab:obsprop}).

The remaining lifetime of a pulsar binary is defined by the shorter
of the merger time of the binary due to the emission of GWs, \tmrg, or the
time that the pulsar will reach the ``death line'', \td~{\citep{rs75}}.
For young pulsars like J1141$-$6545 which have relatively short
radio lifetimes, \td $<$\tmrg. Recycled pulsars, on the other hand,
have far smaller spin-down rates than young pulsars so that
it is likely that close binaries containing a recycled pulsar
will coalesce before the pulsar reaches the death-line (\tmrg $<$\td). 
For circular orbits, \nocite{p64} the results of Peters (1964)
calculations for the merger time of a binary system of two
point masses $m_1$ and $m_2$ with orbital period $P_b$ can be written
as simply:
\begin{equation}
\label{coaltime}
\tau_{\rm mrg}     =  9.83 \times 10^6 \, {\rm yr} \,
                  \left(\frac{P_{b}}{\rm hr}\right)^{8/3}
                  \left(\frac{\mu}{{\rm M}_{\odot}}\right)^{-1}
		  \left(\frac{m_1+m_2}{{\rm M}_{\odot}}\right)^{-2/3},
\end{equation}
where the reduced mass $\mu = m_1 m_2 / (m_1+m_2)$.  For the eccentric
binary J1141--6545, we use the more detailed calculations of Peters
(1964) to calculate $\tau_{\rm mrg}$. Most of observed coalescing
\nswd~binaries as well as the DNS systems have \tmrg $\sim10^{8-9}$ yr.

Our understanding of pulsar emission is rather poor and therefore it
is not clear how to calculate an accurate time associated with the
termination of pulsar emission and hence \td. 
Here we assume the spin-down torque is
dominated by magnetic-dipole radiation with no evolution of the
magnetic field. The surface magnetic field of a neutron star, $B_{\rm
s}$, can be estimated from the current spin period $P$ (s) and spin-down
rate $\dot{P}$ (ss$^{-1}$):
\begin{equation}\label{eq:bp1}
{B_{\rm s}} = 3.2 \times 10^{19} {({P {\dot{P}})^{1/2} }}~~ {\rm G}~.
\end{equation}
{\citet{cr93}} comprehensively discussed the evolution
of a pulsar period based on different magnetic field structures. Their
results are consistent with previous studies {\citep{rs75,v87}}. We
adopt their case C (eq.\ (9) in their paper) according to which the
radio emission terminates when the ``death-period''
\begin{equation}\label{eq:pd}
P_{\rm d}= \Bigl(\frac{B_{\rm s}}{1.4 \times 10^{11} \,{\rm G}}\Bigr)^{7/13}
\, {\rm s}
\end{equation}
is reached. Assuming that the surface magnetic field remains
constant, we can integrate eq.\ (\ref{eq:bp1}) to calculate the time
for the pulsar period to reach $P_{\rm d}$. We find that
\begin{equation}\label{eq:taud}
\tau_{\rm d} = \left(\frac{P_{\rm d}^2 - P^2}{2 P \dot{P}}\right).
\end{equation}
For PSR J1141$-$6545, we use eqs.\ (\ref{eq:pd}) and (\ref{eq:taud}) 
to find the remaining observable lifetime $\tau_{\rm d} \sim 104$ Myr. 
This is significantly less than $\tau_{\rm mrg}$ for this binary system ($\sim
600$ Myr). Including the modest contribution from the characteristic
age of J1141$-$6545, we take the observable lifetime of the binary
system to be \tlife$=$\tc$+$\td$\sim$105 Myr. We note in passing that
{\citet{eb01}} estimated the remaining lifetime of PSR J1141$-$6545 to
be only $\sim$10 Myr. Although no details of their calculation were presented
in their paper, they probably assumed some decay of the magnetic field
which led to their lower value \td~and hence \tlife.

In the cases of the recycled pulsars J0751+1807 and J1757$-$5322,
which have lower magnetic field strengths and hence longer radio
lifetimes, both binaries will coalesce before the pulsars stop
radiating (i.e.~\tmrg $\ll$ \td), so we calculate their lifetime using
\tlife$=$\tsd$+$\tmrg. The estimated lifetimes are 14.3 Gyr
(J0751+1807) and 12.7 Gyr (J1757$-$5322), about two orders of
magnitude longer than for the young \nswd~J1141$-$6545.

\subsection{Probability Density Function of the Galactic \nswd~Coalescence
Rate Estimates} \label{sec:prob}

The basic strategy we use to calculate $P$(\rtot) is described in detail
in paper I. In brief, using a detailed
Monte Carlo simulation, for each observed \nswd~binary, we determine
the fraction of the population that is actually {\it detectable} by
careful modeling of all large-scale pulsar surveys. We include the
selection effects that reduce the detectability of short-period
binary systems when integration times are significant in comparison. 
Our pulsar population model takes into account the distribution
of pulsars in the Galaxy and their luminosity function. Treating each
pulsar seperately, our simulations effectively probe the specific
pulsar sub-populations with pulse and orbital characteristics similar
to those of PSR J0751+1807, J1757$-$5322, or J1141$-$6545.

From the simulations we obtain \nobs~pulsars detected by the
surveys out of a Galactic population of \ntot~in each model.
We calculate \nobs~repeatedly for a fixed \ntot. 
As shown in detail in paper I, the distribution of \nobs~follows a Poisson function,
$P$(\nobs;\nmean). We calculate the best-fit value of \nmean, which is
the mean number of observed pulsars in a given sample, for a given
\ntot. Since we consider each observed pulsar separately, we set 
\nobs~$=1$. For example, one PSR J0751+1807 and no other pulsar similar
to this (in terms of spin and orbital properties of the pulsar) have
been observed. The likelihood of detecting {\it one pulsar} similar
to the observed one from the given pulsar population with
\ntot~samples is simply $P$(1;\nmean). We vary \ntot~and
calculate $P$(1;\nmean) to determine the most probable value of \ntot.
Also, we found \nmean~is directly proportional to \ntot.
We calculate $\alpha$, which is the slope of the function \nmean$=\alpha$\ntot
~for each observed system for a pulsar population model.

Then, using Bayes' theorem, we calculate
$P$(\nmean) from the likelihood $P$(1;\nmean) and eventually calculate
$P$(\rate) using a change of variables. We repeat the whole procedure
for all three observed coalescing \nswd~binaries, and combine the three individual PDFs to 
obtain a total PDF of Galactic coalescence rate of NS-WD binaries, $P$(\rtot). 

In paper I, we showed that a normalized PDF of the coalescence rate for
an individual pulsar binary system can be written as follows:
\begin{equation}\label{eq:pdf1}
P_{\rm i}({\cal R})= C_{\rm i}^{2}~{\cal R}~e^{-C_{\rm i}{\cal R}} ~,
\end{equation}
where $C_{\rm i}$ is a coefficient determined by properties of the
$i^{\rm th}$ pulsar:
 \begin{equation}\label{eq:coef}
C_{\rm i} \equiv {\biggl({ \alpha \tau_{\rm life} 
\over {f_{\rm b}} }\biggr)}_{\rm i}~.
 \end{equation}
Here, the beaming correction factor $f_{\rm b}$ is the inverse of the
fraction of $4 \pi$ sr covered by the pulsar radiation beam during
each rotation. In the case of the two DNS systems, PSRs~B1913+16 and
B1534+12, \citet{k01} adopted $f_b\sim6$ based on pulse profile and
polarization measurements of two pulsars. The lack of such
observations for the current sample of \nswd~binaries means that it is
difficult to estimate reliable values of $f_b$. Therefore, in this
paper, we do not correct for pulsar beaming (i.e.~$f_{\rm b}=$ 1). As
a result, all our values should be considered as lower limits.

In paper I, we calculated $P$(\rtot) considering two observed
DNSs systems (labeled by the subscripts 1 and 2). 
We defined the total rate ${\cal R}_{\rm +} \equiv {\cal R}_{\rm 1} + {\cal R}_{\rm 2}$ and showed that
\begin{equation} \label{eq:pdf2}
P({\cal R}_{\rm +}) = {\Bigl( { C_{\rm 1}C_{\rm 2} \over C_{\rm
2}-C_{\rm 1}}\Bigr)^{2}} \Bigl[ {{\cal R}_{\rm +}} {\Bigl({
e^{-C_{\rm 1}{\cal R}_{\rm +}} + e^{-C_{\rm 2}{\cal R}_{\rm +}}
}\Bigr)} - {\Bigl( {2 \over C_{\rm 2}-C_{\rm 1}} \Bigr)} {\Bigl({
e^{-C_{\rm 1}{\cal R}_{\rm +}} - e^{-C_{\rm 2}{\cal R}_{\rm +}}
}\Bigr)} \Bigr]~,
\end{equation}
where $C_{\rm 1} < C_{\rm 2}$. In Appendix \ref{app:3systems},
we show that this can be extended for the current case of
interest where we have three binary systems such that ${\cal R}_{\rm +} 
\equiv {\cal R}_{\rm 1} + {\cal R}_{\rm 2} + {\cal R}_{\rm 3}$. This
leads to 
\begin{eqnarray} \label{eq:pdf3}
P({\cal R}_{\rm +}) 
&=& \frac{C_{\rm 1}^{2}C_{\rm 2}^{2}C_{\rm 3}^{2}}{(C_{\rm 2}-C_{\rm 1})^{3}(C_{\rm 3}-C_{\rm 1})^{3}(C_{\rm 3}-C_{\rm 2})^{3}} \Biggl[ \\ \nonumber 
           & &  {(C_{\rm 3}-C_{\rm 2})^3}e^{-C_{\rm 1}{\cal R}_{\rm +}}
            \biggl[\ -2(-2C_{\rm 1}+C_{\rm 2}+C_{\rm 3})+{\cal R}_{\rm +}\bigl[-C_{\rm 1}(C_{\rm 2}+C_{\rm 3})+(                   C_{\rm 1}^{2}+C_{\rm 2}C_{\rm 3})\bigr] \biggr]\\ \nonumber
           &+& {(C_{\rm 3}-C_{\rm 1})^{3}}{e^{-C_{\rm 2}{\cal R}_{\rm +}}} 
             \biggl[ 2(C_{\rm 1}-2C_{\rm 2}+C_{\rm 3})+{\cal R}_{\rm +}\bigl[C_{\rm 2}(C_{\rm 3}+C_{\rm 1})-(C_{\rm 2}^{2}+C_{\rm 3}C_{\rm 1})\bigr] \biggr]\\
           &+& {(C_{\rm 2}-C_{\rm 1})^{3}}{e^{-C_{\rm 3}{\cal R}_{\rm +}}} 
             \biggl[ -2(C_{\rm 1}+C_{\rm 2}-2C_{\rm 3})+{\cal R}_{\rm +}\bigl[-C_{\rm 3}(C_{\rm 1}+C_{\rm 2})+(C_{\rm 3}^{2}+C_{\rm 1}C_{\rm 2})\bigr] \biggr] \nonumber
             \Biggr]~,
\end{eqnarray}
where the coefficients $C_{\rm i}$ ($i=1,2,3)$ are defined by eq.\ (\ref{eq:coef}) 
and $C_{\rm 1}<C_{\rm 2}<C_{\rm 3}$.
This result was already used in our recent rate estimation for 
DNS systems to include the newly discovered pulsar, J0737-3039 (Burgay et
al.\ 2003\nocite{burgay03}; Kalogera et al.\ 2004\nocite{k04}). 
As before, the confidence limit
(CL)\footnote{Strictly speaking,
the phrase ``confidence limit'' used in this paper
should be interpreted as a confidence {\it interval} such that
a true value of the Galactic coalescence rate would exist in
a given range of rates. However, we keep the terminology {\it confidence limit}
for consistency with paper I and Kalogera et al.\ (2004).} 
and the lower and upper limits (${\cal R}_{\rm L}$ and ${\cal
R}_{\rm U}$) of the coalescence rate estimates are defined in the same
way we described in paper I, i.e.
\begin{equation} \label{eq:cl}
\int_{{\cal R}_{\rm L}}^{{\cal R}_{\rm U}} P({\cal R}_{\rm +})~
d{\cal R}_{\rm +} = {\rm CL}~,
\end{equation}
and
\begin{equation}
P({\cal R}_{\rm L})=P({\cal R}_{\rm U})~.
\end{equation}

\section{RESULTS}
\label{sec:results}

In Fig.~\ref{fig:prtot}, we show the resulting $P$(\rtot) for
\nswd~binaries along with the individual PDFs for each observed
coalescing binaries. The figure shown here is obtained from our
reference model (model 6 in paper I). As we found for the DNS systems
in paper I, $P$(\rtot) is highly peaked and dominated by a single
object. In this case, PSR J1141$-$6545 dominates the results by virtue
of its short observable lifetime ($\tau_{\rm life} \sim$105 Myr). This
is in spite of the fact that the estimated total number of binaries
similar to PSR J0751+1807 ($N_{\rm 0751}\simeq$ 2900) is the largest among
the observed systems.

We summarize our results for different pulsar population models in
Table \ref{tab:results}. The model parameters are identical to those
described in paper I. We note however, following Kalogera et
al.\ (2004), that our reference model is now model 6 ($L_{\rm min}=$
0.3 mJy kpc$^{2}$) rather than model 1 ($L_{\rm min}=$ 1.0 mJy
kpc$^{2}$). This choice reflects the recent discoveries of faint
pulsar with 1400-MHz radio luminosities below than 1.0 mJy kpc$^{2}$
(Camilo \nocite{cam03} 2003). The peak values of $P$(\rtot) lie in the
range between $\sim0.2-10$ Myr$^{-1}$ where the reference model shows a peak around 4 Myr$^{-1}$. 
For the reference model, the uncertainties in the rates, (defined by ${\cal R}_{\rm U}/{\cal R}_{\rm L}$)
 are estimated to be $\sim$ 6, 27 and 62 at 68\%, 95\%, and 99\% CL, respectively. 
Comparing this to results from {\citet{k04}},
we find that the uncertainties at different CL of the coalescence rate of
\nswd~binaries are typically larger by factor of $\sim$1.4 than those of the DNS systems. 
This result is robust for all models we consider.

The correlations between the peak value of the total Galactic
coalescence rate ${\cal R}_{\rm peak}$ and the model parameters
(e.g. the cut-off luminosity $L_{\rm min}$ and the power index $p$) 
seem to be similar to those of DNSs we observed in paper I. 
As we found for the DNS systems, \rpeak~values strongly
depend on a pulsar luminosity function 
rather than a spatial distribution of pulsars in the Galaxy, in other words, 
\rpeak~values rapidly increase as the fraction of faint pulsars increases.

\section{GRAVITATIONAL WAVE BACKGROUND DUE TO \nswd~BINARIES} 
\label{sec:lisa}

Close binaries consisting of compact objects (e.g. NS$-$WD binaries) 
are suggested as important GW sources in a frequency range below 1 mHz. 
In this range, due to the large number of sources, LISA would not be able 
to resolve each source within a given frequency band. Hence the Galactic 
binaries are expected to establish a confusion noise level (or ``background'') 
dominated by WD$-$WD binaries {\citep{bh90,bh97,nyp01,s01}}. 
In this work, we consider the contribution from \nswd~binaries 
to the predicted confusion noise level. Using our results from the previous section, 
we calculate the amplitude of GW signals from \nswd~binaries in the nearby Universe and compare it with the LISA sensitivity curve\footnote{We use the online sensitivity curve generator to calculate the sensitivity curve of LISA (http://www.srl.caltech.edu/~shane/sensitivity/MakeCurve.html).}. In this work, we assume that the three observed systems represent the whole population of \nswd~binaries in our Galaxy. 

We calculate the characteristic strain amplitude of GWs ($h_{\rm c}$) 
from \nswd~binaries using the results given by Phinney (2001) for circular binaries. 
In general, binaries with an eccentricity $e$ emit GWs at frequencies
$f=n\nu$, i.e.~the $n$-th harmonic of the orbital frequency $\nu$. In
the case of circular binaries, $n=2$ due to the orbital symmetry and the
quadrupole nature of GWs. The eccentricity of PSRs J0751+1807 and
J1757$-$5322 are $e\sim10^{-4}$ and $\sim 10^{-6}$, respectively. 
Hence, it is safe to consider them as circular binaries. 
PSR J1141$-$6545 has an appreciable eccentricity ($e=0.17$),
but for simplicity, we consider only the $n=2$ harmonic as if it were
a circular binary\footnote{We calculate the power distribution in
various harmonics for this eccentricity based on the result of
{\citet{pm63}}. We note that the GW amplitude calculated for
J1141$-$6545 in this work corresponds to $\sim$70\% of the total power
of the gravitational radiation emitted from this binary. The remaining
power is contributed from the higher harmonics.}.

For an observation of length $T_{\rm obs}$ with a GW detector,
the contribution from background sources (\nswd~binaries in this work) 
depends on the number of sources within the frequency resolution, $\Delta f = 1/T_{\rm obs}$. 
Following \citet{s01}, we define an effective GW amplitude 
$h_{\rm rms}(f) \equiv h_{\rm c}(f) {({\Delta f}/ f)}^{1/2}$, where $h_{\rm c}(f)$ is 
a characteristic strain amplitude. {\citet{p01}} showed a simple analytic formula 
to calculate $h_{\rm c}(f)$ for a population of inspiraling circular-orbit binaries 
with a given number density in the nearby Universe. 
We use eq.\ (16) in his paper to calculate 
$h_{\rm c}(f)$\footnote{We assume a case that the last term in Phinney's eq.\ (16), 
${\Bigl(\frac{<{(1+z)^{-1/3}}>}{0.74}\Bigr)}^{1/2}$ becomes unity.
The calculation is not significantly affected by different assumptions on cosmological models and 
comoving number density functions of coalescing binaries (Phinney (2001)).}, and find
 \begin{equation}\label{eq:hrms}
 h_{\rm rms} (f) \simeq 1.7 \times {10^{-26}}
 {\Bigl( {\frac {{\cal M}} {{\rm M}_{\rm \odot}}   } \Bigr)}^{5/6} 
 {\Bigl( {\frac {f} {{\rm mHz}}             } \Bigr)}^{-7/6}
 {\Bigl( {\frac {N_{\rm o}} {{\rm Mpc}^{-3}} } \Bigr)}^{1/2} 
 \Bigl(\frac{T_{\rm obs}}{{{\rm yr}}}\Bigr)^{-1/2}~,
 \end{equation}
where \chmass~is the ``chirp mass'' of a \nswd~binary  
defined by
 \begin{equation}\label{eq:chirp}
 {\cal M}\equiv \frac {(M_{\rm NS} M_{\rm WD})^{3/5}}
                      {(M_{\rm NS} + M_{\rm WD})^{1/5}}~,
 \end{equation}
 and $N_{\rm o}$ is the comoving number density of \nswd, i.e.~the number of sources per Mpc$^{3}$.

We calculate the GW amplitude of \nswd~binaries in a frequency range $f_{\rm min} < f < f_{\rm max}$. 
Estimated GW frequencies of three \nswd~binaries based on their current separations are 
all less than $\sim 0.1$ mHz. In our calculation, however, we set the minimum frequency 
$f_{\rm min}$ to be 1 mHz taking into account the fact that the confusion noise level 
is mainly dominated by Galactic WD--WD binaries at lower frequency range $f<1$mHz {\citep{nyp01}}. 
The maximum GW frequency $f_{\rm max}$ is calculated by $f_{\rm max}=2/P_{\rm b,min}$, 
where $P_{\rm b,min}$ is the minimum orbital period of the binary at the WD Roche-lobe overflow. 

Following {\citet{egg83}}, we calculate the minimum possible separation of 
the binary using $a_{\rm min} = R_{\rm WD}/r_{\rm L}$, where $r_{\rm L}$ is the 
effective Roche lobe radius. We estimate the radius of a white dwarf companion 
$R_{\rm WD}$ adopting the results given by {\citet{tout97}}. 
We show the estimated mass of white dwarf companions in Table 1. 
Converting $a_{\rm min}$ to $f_{\rm max}$ based on the Kepler's 3rd law, we find that 
\begin{equation}\label{eq:fmax}
f_{\rm max} \simeq 0.16~ {r_{\rm L}}^{3/2} 
                   {\Bigl(\frac{M_{\rm WD} + M_{\rm NS}}{M_{\rm \odot}}\Bigr)^{1/2}} 
                   {\Bigl[ 
                          {\Bigl({\frac{M_{\rm ch}}{M_{\rm WD}}}\Bigr)^{2/3} - 
                           \Bigl({\frac{M_{\rm WD}}{M_{\rm ch}}}\Bigr)^{2/3}}
                    \Bigr]^{-3/4}}~~{\rm Hz}~,
\end{equation}
where $M_{\rm ch}=1.44$\solarM~is the Chandrasekhar limit. 
We note that $f_{\rm max}=f_{\rm max} (M_{\rm NS}, M_{\rm WD})$. 
Based on eq.\ (\ref{eq:fmax}), we define three frequency regions: 
(a) $f_{\rm min} < f < f_{\rm max,~0751}$, (b) $f_{\rm
max,~0751} < f < f_{\rm max,~1757}$, and (c) $f_{\rm max,~1757} < f <
f_{\rm max,~1141}$. In the region (a), for example, all three observed
\nswd~systems contribute to the GW background. 
However, for frequencies larger than $f_{\rm max,~0751}$, 
PSR J0751+1807-like populations have already reached the Roche lobe overflow 
and we can not apply eq.\ (\ref{eq:hrms}) to these systems. 
Therefore, in a frequency region (b), 
we consider J1141$-$6545-like and J1757$-$5322-like populations.
Similarly, for the highest frequency range (region (c)), we consider the contribution 
to the GW signals from PSR J1141$-$6545-like population only. 


In order to calculate $h_{\rm rms}$(f),
we need the chirp mass \chmass~and present-day comoving number density of \nswd~binaries $N_{\rm o}$. 
Following {\citet{fp02}}, we define the ``flux-weighted'' averaged chirp mass:
\begin{equation}{\label{eq:chirpave}}
<\cal M> \equiv \frac{\sum F_{\rm gw,i} ~{\cal M_{\rm i}}}{\sum F_{\rm
gw,i}}= \frac{\sum {N_{\rm {peak,i}}~{\cal M_{\rm i}}^{\rm 13/3}}}{\sum
{N_{\rm peak,i}~{\cal M_{\rm i}}^{\rm 10/3}}}~,
\end{equation}
where ${\cal F}_{\rm gw}$ is the GW flux (${\cal F}_{\rm gw} \propto f^{10/3}{\cal M}^{10/3}~{{\cal N}_{\rm peak}}$) and ${\cal N}_{\rm peak}$ is the peak value of $P(N_{\rm tot})$ of the each sub-population of \nswd~binaries in our Galaxy. The subscript $i$ represents each pulsar sub-population. 
For example, in region (b), $<$${\cal M}$$>=({\cal N}_{\rm 1757}{{\cal M}_{\rm 1757}}^{13/3}+{\cal N}_{\rm 1141}{{\cal M}_{\rm 1141}}^{13/3})/({\cal N}_{\rm 1757}{{\cal M}_{\rm 1757}}^{10/3}+{\cal N}_{\rm 1141}{{\cal M}_{\rm 1141}}^{10/3})$. 
Because ${\cal N_{\rm peak}}$~is a constant and independent of the GW frequency, 
it follows that $<$$\cal M$$>$ is independent of frequency. 
As a result, the evolution of orbital characteristics and hence the GW frequency of a binary 
are solely determined by the inspiral process. 
This is true regardless of the inital distribution of orbital characteristics.

We now calculate the comoving number density of \nswd~binaries
\begin{equation}
 N_{\rm o} = \int_{0}^{\infty} N(z) dz~,
\end{equation}
where $N(z)dz$ is the number of \nswd~binaries per unit comoving
volume between redshift $z$ and $z+dz$. Noting that the number density of
\nswd~binaries is proportional to the total number of systems, 
we may write
\begin{equation}\label{eq:no}
 N_{\rm o} =\epsilon~{\cal N}_{\rm PSR}~,
\end{equation}
where $\epsilon$ is the star formation rate density 
per unit comoving volume ($\dot\rho$) normalized to the Galactic 
star formation rate ($r$) i.e.~$\epsilon = \dot{\rho}/r$ (in Mpc$^{-3}$). 
${\cal N}_{\rm PSR}$ is the most likely value of the total number of pulsars for each frequency range. 
(e.g. ${\cal N}_{\rm PSR}={\cal N_{\rm 0751}}+{\cal N_{\rm 1757}}+{\cal N_{\rm 1141}}$ in region (a)). 
We derived $P(N_{\rm tot})$ for individual systems in paper I. 
In a similar fashion to the coalescence rate estimation described in 
Appendix \ref{app:3systems}, the combined PDF of \ntot~can be calculated 
from individual PDFs of the observed systems. 
Then ${\cal N}_{\rm PSR}$ can be obtained from the peak value of the PDF 
we calculate for each frequency range we discussed above. 

Using the results of Cole et al.\ (2001){\nocite{cole01}}, we find
\begin{equation}\label{eq:cole2}
 \epsilon = {\frac{\dot{\rho}(0)}{r}}~\int_{0}^{z_{\rm max}} \frac{(1 + {\frac{b}{a}}z)}{(1+({\frac{z}{c}})^{d} )}dz~~{\rm Mpc}^{-3}~,
\end{equation}
where $(a, b, c, d)=(0.0166, 0.1848, 1.9474, 2.6316)$ are parameters
which take into account dust-extinction corrections (see \citet{cole01} for further details).
Assuming a Hubble constant H$_{\rm o}=65$ km s$^{-1}$ Mpc$^{-1}$, 
we calculate the Galactic star formation rate density $\dot{\rho}(0)\simeq0.01$ \solarM~yr$^{-1}$ Mpc$^{-3}$. 
Following {\citet{cet99}}, assuming the Salpeter
initial mass function, we convert the Galactic supernova type SN$_{\rm II+Ib/c}$
rate\footnote{{\citet{cet99}} assumed a Hubble constant H$_{\rm o}=75$
km s$^{-1}$ Mpc$^{-1}$. Since we adopt H$_{\rm o}=65$ km s$^{-1}$
Mpc$^{-1}$, we have multiplied their results by a factor (65/75)$^2$.
We also note that we consider Sbc-Sd type galaxies only, which would
be relevant for active star-forming regions.}
to the star formation rate finding $r\sim 0.7$ \solarM yr$^{-1}$. 
Numerically integrating eq.~(\ref{eq:cole2})
out to $z_{\rm max}=5$, which is considered to be the onset of the galaxy formation {\citep{s01}}, 
we find $\epsilon \sim 0.6$ Mpc$^{-3}$. The number density of \nswd~binaries $N_{\rm o}$ then can
be calculated by eq.\ (\ref{eq:no}) for a given ${\cal N}_{\rm PSR}$ for each model.

In Fig.~\ref{fig:lisa}, we plot the GW amplitude $h_{\rm rms}$ 
against the simulated LISA sensitivity curve calculated for a signal-to-noise 
ratio S/N=1. 
%
All dotted lines correspond to the GW amplitude calculated from 
the full set of pulsar population models we consider and the solid line 
is the result from our reference model. The range of the GW amplitude 
for all models spans about an order of magnitude. 
We find that the GW background amplitude from \nswd~binaries is about 1--2 orders 
of magnitude smaller than the expected sensitivity curve of LISA at GW frequencies 
larger than 1mHz and it is unlikely that this population will be detected with LISA. 
In the lower frequency region (below $\sim$ 1mHz), the GW background amplitude 
from \nswd~binaries increases as $f$ decreases. However, the contribution from \nswd~binaries 
to the GW background noise level would still be less than $\sim$ 10\% of 
the GW amplitude from WD--WD binaries (dashed line in \ref{fig:lisa}). 
We note that, however, we have not considered any beaming corrections, 
so the \nswd~background curves should be viewed as lower limits. 
This possibility is discussed briefly in the next section.

\section{DISCUSSION} \label{sec:discussion}

We have used detailed Monte Carlo simulations to calculate the
Galactic coalescence rate of \nswd~binaries. From the reference
model, the most probable value of ${\cal R}_{\rm tot}$ is estimated to
be $4.11_{-2.56}^{+5.25}$ Myr$^{-1}$ at a 68\% statistical confidence
limit. We find that the coalescence rate of \nswd~binaries is about
factor of 20 smaller than those of DNS for all pulsar population models
we consider. As mentioned above, we did not take into account any
beaming correction for \nswd~binaries. If we assume a beaming
fraction of pulsars in \nswd~binaries similar to that of pulsars found
in DNS, $f_{\rm b}\sim6$, then the discrepancy beween \rpeak~(DNS) and
\rpeak~(\nswd) is signicantly reduced.
As a simple estimate, if we assume $f_{\rm b,1141}\sim5$, but keeping
$f_{\rm b}=1$ for the other two binaries, the estimated Galactic
coalescence rate increases to $18.06_{-12.74}^{+26.05}$ Myr$^{-1}$ 
at a 68\% confidence limit.
Hence the ratio between DNS and \nswd~coalescence rate
decreases to about 5. Because the contribution from PSRs J0751+1807
and J1757--5322 is an order of magnitude smaller than that of
J1141--6545, moderate values of beaming fraction for those recycled
pulsars do not change the result significantly. 

Based on the number of sources of \nswd~binaries in our Galaxy, 
we estimate the effective GW amplitude from the cosmic population of these systems. 
We find that the GW background from \nswd~binaries is too weak to be detected by LISA 
for the nominal beaming correction. Only by adopting an unreasonably large beaming 
correction factor, $f_{\rm b}>10$, could these systems be detectable by LISA in the mHz range. 
These results are in good agreement with an independent study by Cooray (2004) 
based on statistics of low mass X-ray binaries.

We finally note that combining the results from paper I and this work
can give us strong constraints on the population synthesis models.
The preferred models, which are consistent with both 
${\cal R_{\rm NS-WD}}$ and ${\cal R_{\rm DNS}}$, can then be used for the
estimation of the coalescence rate of neutron star -- black hole
binaries, which have not yet been observed.

\acknowledgments
We thank A. Cooray for noticing the frequency resolution 
correction factor in the calculation of the GW background.
This work is partially supported by NSF grant PHY-0121420 and a
Packard Fellowship in Science and Engineering to VK. DRL is a
University Research Fellow funded by the Royal Society. He is also
grateful for support from the Theoretical Astrophysics Visitors' Fund
at Northwestern University.

\appendix
\section{Combined rate PDF for three binary systems}
\label{app:3systems}
In paper I, we derived expressions for $P$(\rtot) for one
and two coalescing binaries.
In this paper, and in our revised DNS coalescence rate estimates
(Kalogera et al.~2004), we extend this PDF to the case of three systems.
Following paper I, we define a coefficient for each observed \nswd:
\begin{equation}
A \equiv \Bigl( \frac{\alpha {\tau_{\rm life}}} {f_{\rm b}}\Bigr)_{\rm 1141},
B \equiv \Bigl( \frac{\alpha {\tau_{\rm life}}} {f_{\rm b}}\Bigr)_{\rm 0751}~, ~{\rm and}~
C \equiv \Bigl( \frac{\alpha {\tau_{\rm life}}} {f_{\rm b}}\Bigr)_{\rm 1757}~,
\end{equation}
where $A<B<C$. Recall that $\alpha$ is the slope of the function
\nmean$=\alpha$\ntot~and is determined for each pulsar population
model for each \nswd~system. By definition, the total Galactic
coalescence rate is the sum of all three observed systems:
\begin{equation}
{\cal R}_{\rm +} \equiv {\cal R}_{\rm 1} + {\cal R}_{\rm 2} +{\cal R}_{\rm 3}~.
\end{equation}
Redefining ${\cal R}_{\rm +} \equiv {\cal R}_{\rm a} + {\cal R}_{\rm
b}$, where ${\cal R}_{\rm a} \equiv {\cal R}_{\rm 1} + {\cal R}_{\rm
2}$ and ${\cal R}_{\rm b} \equiv {\cal R}_{\rm 3}$, we transform
${\cal R}_{\rm a}$ and ${\cal R}_{\rm b}$ to new variables ${\cal
R}_{\rm +}$ and ${\cal R}_{\rm -} \equiv {\cal R}_{\rm a} - {\cal
R}_{\rm b}$
\begin{equation}
 P({\cal R}_{\rm +}, {\cal R}_{\rm -}) 
= P({\cal R}_{\rm a},{\cal R}_{\rm b}) \left \vert \matrix {
{{d{\cal R}_{\rm a}} \over {d{\cal R}_{\rm +}} } & {{d{\cal R}_{\rm b}} \over {d{\cal R}_{\rm -}} } \cr
{d{\cal R}_{\rm a} \over d{\cal R}_{\rm -} } & {d{\cal R}_{\rm b} \over d{\cal R}_{\rm +}} \cr } \right
\vert = {1 \over 2} P({\cal R}_{\rm a},{\cal R}_{\rm b})~.
 \end{equation}
Since both ${\cal R}_{\rm +}$ and ${\cal R}_{\rm -}$ are positive, 
$-{\cal R}_{\rm -}\leq {\cal R}_{\rm +} \leq + {\cal R}_{\rm -}$. 

The PDF of the total rate ${\cal R}_{\rm +}$ is obtained after
integrating $P({\cal R}_{\rm +}, {\cal R}_{\rm -})$ over ${\cal
R}_{\rm -}$:
\begin{equation}\label{eq:ptotint}
 P({\cal R}_{\rm +}) = \int_{\rm {\cal R}_{\rm -}} P({\cal R}_{\rm +},{\cal R}_{\rm -}) \, d{\cal R}_{\rm -} =
 {1 \over 2} \int_{\rm {\cal R}_{\rm -}} P({\cal R}_{\rm a}, {\cal R}_{\rm b}) \, d{\cal R}_{\rm -}~,
 \end{equation}
where 
 \begin{equation}
P({\cal R}_{\rm a},{\cal R}_{\rm b}) = P({\cal R}_{\rm a})P({\cal R}_{\rm b})~.
 \end{equation}
Here, $P({\cal R}_{\rm a}= {\cal R}_{\rm 1} + {\cal R}_{\rm 2})$ is
given by eq.\ (\ref{eq:pdf2}) and we can rewrite the formula with
appropriate coefficients defined earlier:
\begin{equation}\label{eq:pdf2a} 
P({\cal R}_{\rm a}) = {\Bigl( { AB \over B-A}\Bigr)^{2}} \Bigl[ {{\cal
R}_{\rm a}} {\Bigl({ e^{-A{\cal R}_{\rm a}} + e^{-B{\cal R}_{\rm a}}
}\Bigr)} - {\Bigl( {2 \over B-A} \Bigr)} {\Bigl({ e^{-A{\cal R_{\rm a}}} - e^{-B{\cal R}_{\rm a}} }\Bigr)} \Bigr]~.
\end{equation}
The individual PDF $P({\cal R}_{\rm b}) \equiv P({\cal R}_{\rm 3})$
(eq.\ (\ref{eq:pdf1})) is also rewritten as follows:
\begin{equation}\label{eq:pdf1a}
P_{\rm i}({\cal R}_{\rm b})= C^{2}~{\cal R_{\rm b}}~e^{-C{\cal R}_{\rm b}} ~.
\end{equation}
Replacing ${\cal R}_{\rm a}=(1/2)({\cal R}_{\rm +}+{\cal R}_{\rm -})$
and ${\cal R}_{\rm b}=(1/2)({\cal R}_{\rm +} - {\cal R}_{\rm -})$ in
eq.\ (\ref{eq:ptotint}), the normalized $P({\cal R}_{\rm +})$ can be
obtained by integration. After some algebra, we find:
\begin{eqnarray} \label{eq:pdf3a}
P({\cal R}_{\rm +}) &=& \frac{A^{2}B^{2}C^{2}}{(B-A)^{3}(C-A)^{3}(C-B)^{3}} \\ \nonumber
           & &  \Biggl[ {(C-B)^3}e^{-A{\cal R}_{\rm +}}
            \biggl[\ -2(-2A+B+C)+{\cal R}_{\rm +}\bigl[-A(B+C)+(A^{2}+BC)\bigr] \biggr]\\ \nonumber
                  &+& {(C-A)^{3}}{e^{-B{\cal R}_{\rm +}}} 
             \biggl[ 2(A-2B+C)+{\cal R}_{\rm +}\bigl[B(C+A)-(B^{2}+CA)\bigr] \biggr]\\
                  &+& {(B-A)^{3}}{e^{-C{\cal R}_{\rm +}}} 
             \biggl[ -2(A+B-2C)+{\cal R}_{\rm +}\bigl[-C(A+B)+(C^{2}+AB)\bigr] \biggr]
             \Biggr]~ \nonumber. 
\end{eqnarray}

\clearpage
\begin{deluxetable}{lrrrrcccccrc}
\tablecolumns{12}
\tablewidth{0pc}
\tablecaption{\label{tab:obsprop} Observational properties of NS--WD binaries.
From left to right, the columns indicate the pulsar name, 
spin period $P$, spin-down rate $\dot{P}$, 
orbital period $P_b$, most
probable mass of the WD companion $m_{\rm wd}$, orbital eccentricity $e$,
characteristic age $\tau_c$, spin-down age $\tau_{\rm sd}$, GW
merger timescale $\tau_{\rm mrg}$, time to reach the death line
$\tau_d$, most probable number of NS--WD systems of this type
in the Galaxy and references to this system in the literature.}
\tabletypesize{\footnotesize}
\tablehead{
 
\colhead{PSRs} &
\colhead{$P$} &
\colhead{{$\dot{P}$}} &
\colhead{$P_{\rm b}$ }&
\colhead{m$_{\rm wd}^{a}$} &
\colhead{e} &
\colhead{$\tau_{\rm c}$} &
\colhead{$\tau_{\rm sd}$} &
\colhead{$\tau_{\rm mrg}$} &
\colhead{$\tau_{\rm d}$} &
\colhead{${\cal N}_{\rm PSR}$} &
\colhead{References$^{b}$}
 \\ 
 
\colhead{} &  \colhead{(ms)} &  \colhead{(s~s$^{-1}$)} &  \colhead{(hr)} & \colhead{(M$_{\rm \odot}$)} &
\colhead{} & \colhead{(Gyr)} & \colhead{(Gyr)} &  \colhead{(Gyr)} & \colhead{(Gyr)} & \colhead{} & \colhead{}
}

\startdata

J0751+1807  & 3.479 & 8.08$\times 10^{-21}$  & 6.315  &  0.18 & $<10^{-4}$  & 6.8 & 6.7 & 7.6 & ... & 2900 & 1, 2\\
J1757$-$5322 & 8.870 & 2.78$\times 10^{-20}$ & 10.88 & 0.67 & $10^{-6}$ & 5.1 & 4.9 & 7.8 & ... &1200 & 3\\
\\
J1141$-$6545 & 393.9 & 4.29$\times 10^{-15}$& 4.744 & 0.986  & 0.172 & 1.5$\times 10^{-3}$ & ... & 0.6 & 0.104 &400 & 4,5\\
\enddata
 
\tablenotetext{a}{The assumed NS mass and inclination angle $i$ are 2.2M$_{\rm \sun}$ and $78^{\circ}$ for J0751+1807 (Nice et al. (2004)) and 1.35M$_{\rm \sun}$ and $60^{\circ}$ for J1757$-$5322. For PSR J1141$-$6545, we adopt values given in Bailes et al. (2003).}
\tablenotetext{b}{References: 
(1) Lundgren, Zepka, \& Cordes (1995); (2) Nice, Splaver, \& Stairs (2004)
(3) Edwards, \& Bailes (2001) ; 
(4) Kaspi et al.\ (2000) ; 
(5) Bailes et al.\ (2003).
}

\end{deluxetable}


 
 \begin{deluxetable}{lrllrll}

 \tablewidth{0pc}
 \tabletypesize{\small}
 \tablecaption{\label{tab:results}Estimates for the Galactic coalescence rate (${\cal R}_{\rm tot}$) of NS--WD binaries at various confidence limits for all models considered.}
 \tablehead{
\colhead{Model$^{a}$} & \multicolumn{3}{c}{${\cal R}_{\rm tot}$ (Myr$^{-1}$)}\\
 
\cline{1-4}
 
\colhead{} 
&\colhead{peak$^{b}$} &\colhead{68\%$^{c}$} &\colhead{95\%$^{c}$} 
}

\startdata
1  & 1.23 &$_{-0.77}^{+1.57}$ & $_{-1.04}^{+3.97}$  \\ 
2  & 1.00 &$_{-0.62}^{+1.27}$ & $_{-0.83}^{+3.20}$   \\
3  & 1.32 &$_{-0.83}^{+1.69}$ & $_{-1.12}^{+4.27}$   \\
4  & 1.53 &$_{-0.97}^{+1.99}$ & $_{-1.31}^{+5.02}$   \\
5  & 1.19 &$_{-0.75}^{+1.53}$ & $_{-1.01}^{+3.87}$   \\
\bf{6} & 4.11 &$_{-2.56}^{+5.25}$ & $_{-3.47}^{+13.23}$\\ 
7  & 1.73 &$_{-1.08}^{+2.21}$ & $_{-1.46}^{+5.57}$  \\
8  & 0.83 &$_{-0.52}^{+1.07}$ & $_{-0.71}^{+2.71}$   \\
9  & 0.43 &$_{-0.27}^{+0.55}$ & $_{-0.36}^{+1.39}$   \\
\\
10 & 1.42 &$_{-0.92}^{+1.88}$ & $_{-1.23}^{+4.76}$  \\
11 & 0.72 &$_{-0.47}^{+0.96}$ & $_{-0.62}^{+2.42}$   \\
12 & 0.55 &$_{-0.35}^{+0.72}$ & $_{-0.47}^{+1.82}$   \\
13 & 0.40 &$_{-0.26}^{+0.53}$ & $_{-0.34}^{+1.33}$   \\
14 & 0.23 &$_{-0.15}^{+0.30}$ & $_{-0.20}^{+0.76}$   \\
\\
15 & 10.03&$_{-6.09}^{+12.46}$ & $_{-8.34}^{+31.38}$\\
16 & 3.68 &$_{-2.25}^{+4.57}$ & $_{-3.07}^{+11.53}$  \\
17 & 2.45 &$_{-1.50}^{+3.06}$ & $_{-2.04}^{+7.71}$  \\
18 & 1.55 &$_{-0.95}^{+1.94}$ & $_{-1.29}^{+4.90}$  \\
19 & 0.72 &$_{-0.44}^{+0.91}$ & $_{-0.60}^{+2.29}$  \\
\\
20 & 5.48 &$_{-3.29}^{+6.69}$& $_{-4.52}^{+16.86}$ \\
\\
21 & 1.96 &$_{-1.17}^{+2.37}$ & $_{-1.61}^{+5.96}$ \\
22 & 1.55 &$_{-0.96}^{+1.97}$ & $_{-1.3}^{+4.96}$ \\
23 & 1.13 &$_{-0.71}^{+1.47}$ & $_{-0.97}^{+3.72}$  \\
24 & 1.14 &$_{-0.73}^{+1.49}$ & $_{-0.98}^{+3.76}$  \\
25 & 1.21 &$_{-0.76}^{+1.56}$ & $_{-1.03}^{+3.95}$  \\
26 & 1.33 &$_{-0.84}^{+1.72}$ & $_{-1.13}^{+4.35}$  \\
27 & 1.46 &$_{-0.93}^{+1.90}$ & $_{-1.25}^{+4.79}$  \\
 
\enddata
\tablenotetext{a}{Model No. (see paper I for details of model parameters.)}
\tablenotetext{b}{Peak value of the probability density function.}
\tablenotetext{c}{Confidence limits.}
\end{deluxetable}

\clearpage

\centerline{\psfig{figure=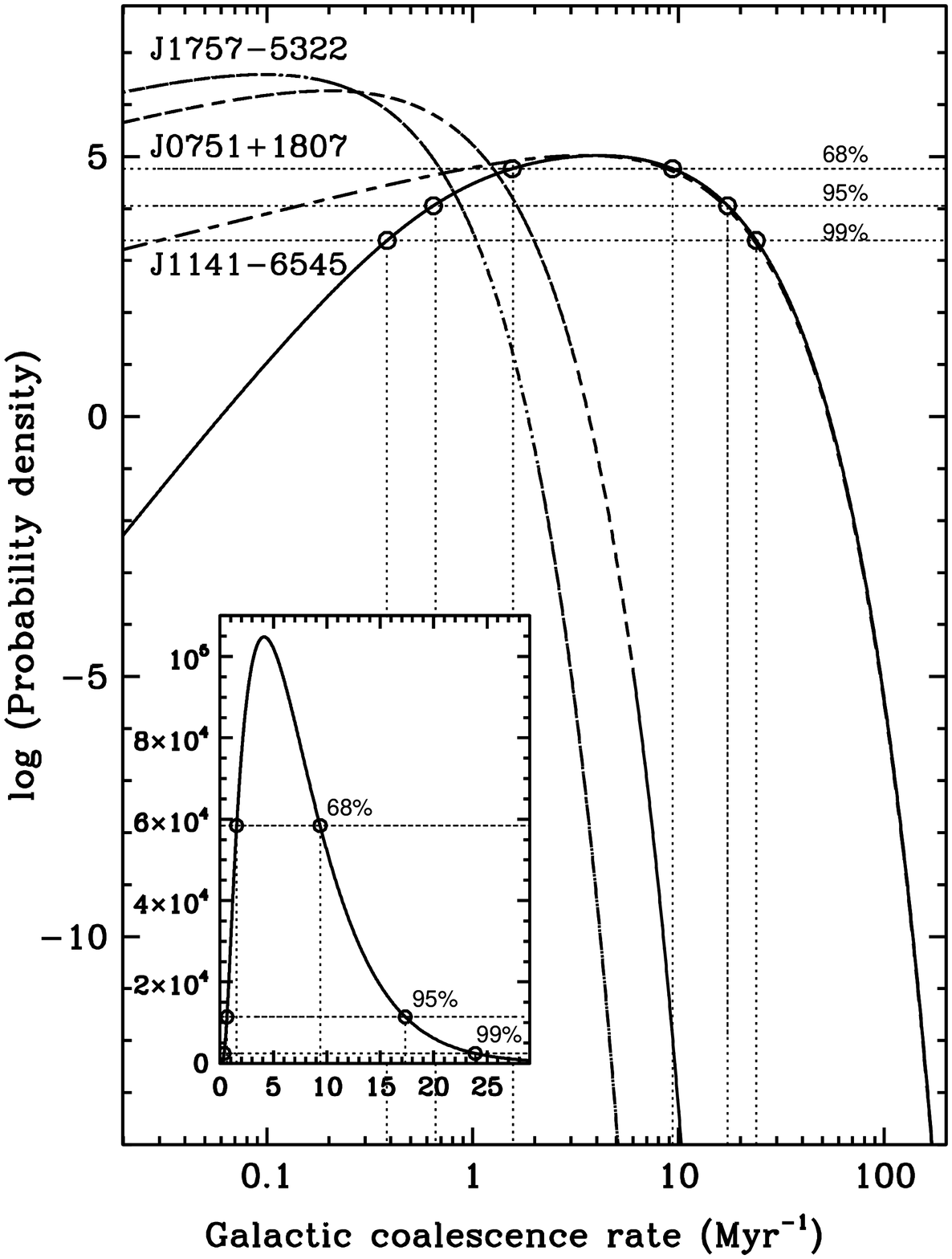,angle=0,width=5.in}} 
\figcaption{The PDFs of the Galactic coalescence rate estimation in
both a logarithmic and a linear scale (inset) are shown for the
reference model. The solid line represents $P({\cal R}_{\rm
tot})$. Other curves are $P({\cal R})$ for PSRs~J1757$-$5322
(dot-dash), J0751+1807 (short-dash), and J1141$-$6545 (long-and-short
dash)-like populations, respectively. Dotted lines correspond to 68\%,
95\%, and 99\% confidence limits for $P({\cal R}_{\rm tot})$.
\label{fig:prtot}}

 \centerline{\psfig{figure=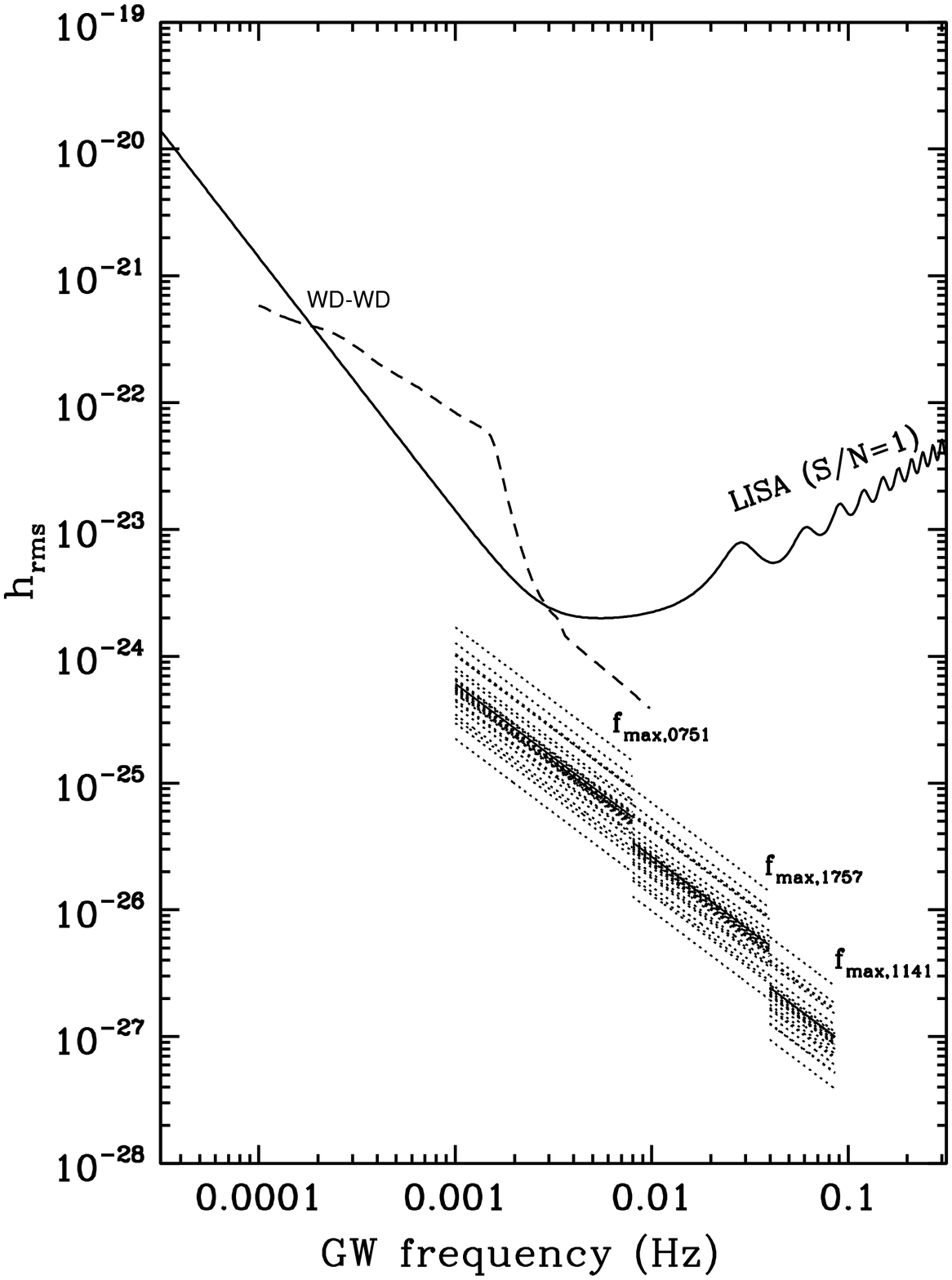,angle=0,width=5.in}} 
\figcaption{The effective GW amplitude $h_{\rm rms}$ for coalescing \nswd~binaries overlapped
with the LISA sensitivity curve. The curve is produced with the
assumption of S/N=1 for 1 yr of integration. Dotted
lines are results from all models we consider except the reference
model, which is shown as a solid line. We also show the expected confusion noise
from Galactic WD--WD binaries for comparison (dashed line). 
\label{fig:lisa}}

\end{document}